\documentclass[12pt,journal,onecolumn,draftcls]{IEEEtran}
\usepackage{graphicx,psfrag,amsmath,amsfonts,verbatim}
\usepackage[small,bf]{caption}
\usepackage{amsmath,bm}
\usepackage[square, comma, sort&compress, numbers]{natbib}
\usepackage{float}
\usepackage{algorithm}
\usepackage{algorithmic}

\usepackage{rotating}
\usepackage{multirow}
\usepackage{slashbox}
\usepackage{booktabs}
\usepackage[ bookmarksopen=true,
             pdfborder=001,
             bookmarksnumbered=true,
          ]{hyperref}

\newcommand{\Expect}{\mathop{\bf E{}}}

\newcommand{\Ei}{{\mathrm{Ei} }}

\newcommand{\ie}{{ i.e.}}

\newcommand{\etal}{{\it et al. }}

\title{An Inter-Node Interference Suppression Approach in Full-Duplex Wireless Communications}
\author{Fei~Wu,  Mintao Zhang, Si Li and Wanzhi MA
\thanks{ Mintao Zhang  are with Xihua University, Town of Hongguang, Pixian, Chengdu, Sichan, 610039. E-mail: zhangmt@mail.xhu.edu.cn. }
\thanks{Fei~Wu, Si Li and~Wanzhi MA are with University of Electronic Science and Technology of China.}
}

\begin{document}

\maketitle

\begin{abstract}
\addcontentsline{toc}{section}{Abstract}
Considering that a full-duplex network is comprised of a full-duplex (FD) base station (BS) and two half-duplex (HD)  users, one  user transmits  on the uplink channel and the other receives through the downlink channel on the same frequency.
The uplink user will generate inter-user interference (IUI) on the downlink user through the interference channel. In this paper, we propose a novel IUI suppression approach when the BS knows the full channel station information. The main idea of the approach is to retransmit the weighted uplink signal as soon as it has been received at the BS.
For the narrowband case, we first derive the closed-form expression of the optimum weighted coefficient when the SI is perfectly cancelled at the BS and then analyze the performance of the proposed IUI suppression approach in practical considerations.
Furthermore, the proposed IUI suppression approach can be extended to the broadband case using a time-domain weighted filter.
Simulation results shows the advantage over existing IUI suppression schemes.
\end{abstract}

\begin{IEEEkeywords}
Full-duplex, inter-user interference suppression, imperfect channel information, achievable rate, energy efficiency.
\end{IEEEkeywords}

\section{Introduction}

\IEEEPARstart{F}{ull-duplex} (FD)  wireless communications simultaneously operate over the same frequency channel which have the potential to double the spectrum efficiency \cite{duarte2010fullduplex,choi2010achieving,radunovic2010rethinking,bharadia2013full}.
 The FD communications are in contrast to the half-duplex (HD) communications, which are either time-division, frequency-division, or code-division for transmitting and receiving.
 The main obstacle in implementing the FD transceiver is the large self-interference (SI) leaking from the local transmitter because of the closeness of the transmitter and receiver chains. Typically, the SI signal is million times stronger than the intended signal on the air.
 Recently, many studies have implemented the FD communication systems using combing SI suppression techniques such as passive suppression, analog suppression and digital suppression and thus the SI can be attenuated to detect the intended signal \cite{riihonen2011mitigation,jain2011practical,duarte2012experimentdriver,sabbarwal2014inband,kim2015asurvey,zhang2016fullduplex,zhang2015impro,Wu2016nearfield}.
The feasibility of FD creates new design opportunities for wireless communications.
Since the SI suppression techniques are too complex for mobile users compared with the base station (BS) to implement in near future,  we focus on studying a network comprising the HD mobile users and the FD BSs.

\begin{figure}
\begin{center}
\includegraphics[width=0.35\textwidth]{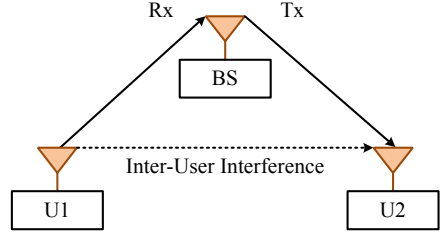}
\end{center}
\centering\caption{FD network with a BS and two users  .}
\label{f-Intern-network}
\end{figure}

Considering about a FD network shown in Fig. \ref{f-Intern-network}, two HD mobile users communicate with a FD BS that supports one  uplink and one downlink traffics at the same time on the same frequency. In this network, there are two types of interference: SI at the FD BS, and inter-user interference (IUI) from the uplink mobile user U1 to the downlink mobile user U2. The distinction of these two types of interferences is that, the SI signal is known at the BS receiver, because the transmitter and receiver are deployed at the same location, but the IUI is not known at the node U2. In this paper, we assume that the SI at the BS has been suppressed under the noise floor and can be neglected, and thus focus on IUI suppression. Note that SI may be easily processed \cite{choi2010achieving,duarte2010fullduplex,radunovic2010rethinking,riihonen2011mitigation,jain2011practical,duarte2012experimentdriver,bharadia2013full,sabbarwal2014inband,kim2015asurvey,zhang2016fullduplex,Syrjl2016analysis,Wu2016nearfield} as the SI information is locally available at the FD BS.
However, handling IUI  is much more challenging as it is between distributed users, who cannot share data information without sacrificing bandwidth resources. Roughly, IUI management techniques can be sorted into three categories as follows

\emph{1) Resource allocation techniques:} To reduce IUI, Goyal \etal propose a scheduling approach to maximize the achievable rate that can be harvested from in-band FD transmissions \cite{goyal2013analyzing,goyal2014improving}.  Ramirez \etal propose a joint algorithm
to realize power allocation and routing considering both SI and IUI among neighboring nodes in FD wireless relay networks \cite{ramirez2013optimal}. The authors \cite{yin2013fullduplex} investigate IUI problem that occurs in multi-user scenarios and show that FD transmission can be made more robust against IUI, which inevitably occurs in cellular communication systems. Shao \etal  \cite{shao2014analysis} propose partitioning method that the cell is divided into several partitions where the IUI is regarded as Gaussian noise at the mobile user receiver and the same frequency resource is assigned  to the two users who are far enough from each other. This method is suitable for the larger cell.  In \cite{nam2015radioresource}, the authors investigate the joint issue of subcarrier assignment and power allocation
to maximize the sum achievable rate performance in FD orthogonal frequency division multiple access (OFDMA) networks. In  \cite{duarte2016interuser} an IUI coordination approach based on geographical context information is given, which exploits the signal attenuation from obstacles between mobile users so that IUI is minimized. To potentially cancel co-channel interference caused by other users, the opportunistic interference suppression (OIC) technique is applied at user side and  a joint mode selection, user scheduling, and channel allocation problem is formulated to maximize the system throughput \cite{yu2016jointuser}.

\emph{2) Medium access control (MAC) techniques:}  The simplest method avoids IUI by picking nodes that are completely hidden from each other \cite{sahai2011pushing}. Singh \etal propose a distributed MAC
protocol with a selection scheme for a secondary receiver \cite{singh2011efficient}. The selection approach allocates different  weight values to candidates for the secondary receiver. If a candidate node experiences more
successful transmissions, it has a higher weight value and thus a greater chance to be selected as the secondary receiver.  Another method \cite{kim2013janus} optimizes user pairing by considering IUI based on the information about IUI and traffic demands reported from all pairs of users.
Goyal \etal in \cite{goyal2013adistributed}  develop a centralized MAC protocol considering  interference between two users due to their concurrent transmissions. In \cite{choi2015powercontrolled}, Choi \etal studies a random-access MAC protocol using distributed power control to manage IUI in wireless networks with FD BSs that serve HD users. In practice, it would be favorable to design an adaptable MAC protocol  configured by specific channel and network conditions.
Chen \etal \cite{chen2015probabilistic} present a distributed FD MAC protocol that allows a BS to adaptively switch between FD and HD modes so as to reduce the influence of the IUI.

\emph{3) Physic layer (PHY) techniques:}  Bai \etal \cite{bai2012decode,bai2013distributed} present the sum achievable rate of a three-node wireless network with a FD BS and two HD terminals. Their method is the first study to access the direct mitigation of IUI among HD terminals and the authors utilize an additional  side-channel to manage IUI. To investigate
the performance of the proposed approach, four  schemes are proposed under different  side-channel information.
To combat the severe IUI in the FD communication systems, Sahai \etal \cite{sahai2013onuplink} propose new interference management
strategies which allow the network to handle IUI while obtaining rate gains by operating in-band FD transmission. In their following work, Sahai \etal  propose an interference management (IA) scheme to handle IUI for multiple antennas FD communication systems so as to achieve rate gains over conventional cellular systems in terms of degrees of freedom (DoF) \cite{sahai2014ondefrees}. Using the cascade interference suppression, IUI can be effectively eliminated among multiple nodes.
In \cite{bi2015onrate}, successive interference suppression (SIC) is applied, which is based on the fact that the downlink mobile user observes a MAC of two users and thus the downlink user has an opportunity to remove the IUI according to the transmission rates and its received powers.
 In their following work \cite{bi2016superposition}, superposition coding based IUI suppression (SCIIC)  referenced by the interference suppression method applied in the X-interference channel is proposed.
  Mai \etal \cite{mai2016defrees} manage the IUI through transmit beamforming in millimeter wave FD systems when the BS adopts multiple antenna.

In the following, we study how to reduce the IUI.
If the BS knows the full channel information of the uplink, the downlink, and the interference channels,   the BS knows the IUI as soon as it receives the uplink signal and thus IUI can be  suppressed  by transmitting the reversed signal of IUI at the BS. This idea is simple.
For example, we assume the values of the uplink, the downlink, and the interference channels are $1$s and the uplink signal is $x$.
When  signal-to-noise (SNR) is high and noise can be neglected, the IUI signal is $x$ and the IUI signal can be perfectly suppressed if the BS transmits a superposition signal $-x$. This scheme does not need complex algorithm and only need some extra transmit power at the BS.
 Our contributions are summarized as follows:
 1), first, we propose a BS assisting IUI suppression scheme and give the closed-form expression for narrowband system when the SI is perfectly cancelled at the BS;
 2), second, the performance of the proposed IUI suppression scheme is furthermore evaluated under practical considerations such as imperfect SI mitigation at the base station, imperfect channel information, delay difference between the IUI and the IUI suppression signals, power control and limited total transmit power;
 3), third,  besides the newest IUI suppression scheme,  we compared the performance of the proposed IUI suppression scheme with the HD mode,  the ideal FD mode without IUI, and the FD mode with IUI  but not suppressed;
 4), finally, we extend the proposed approach to the broadband case.

The remainder of this paper is organized as follows.
In Section II, the proposed IUI suppression scheme is given.
The performance of the proposed scheme for narrowband system is introduced in Section III.
Section IV extends the proposed scheme to the broadband system. Section V  presents the conclusion.

\emph{Notation:} $\Expect$ denotes the expectation operation, $^\ast$ denotes the complex conjugate, and $\mathcal{CN} \left(\mu, \delta^2 \right)$ denotes circularly symmetric Gaussian random variable with $\mu$ mean and $\delta^2$ variance.

 \section{Proposed IUI suppression scheme}

\begin{figure}
\begin{center}
\includegraphics[width=0.35\textwidth]{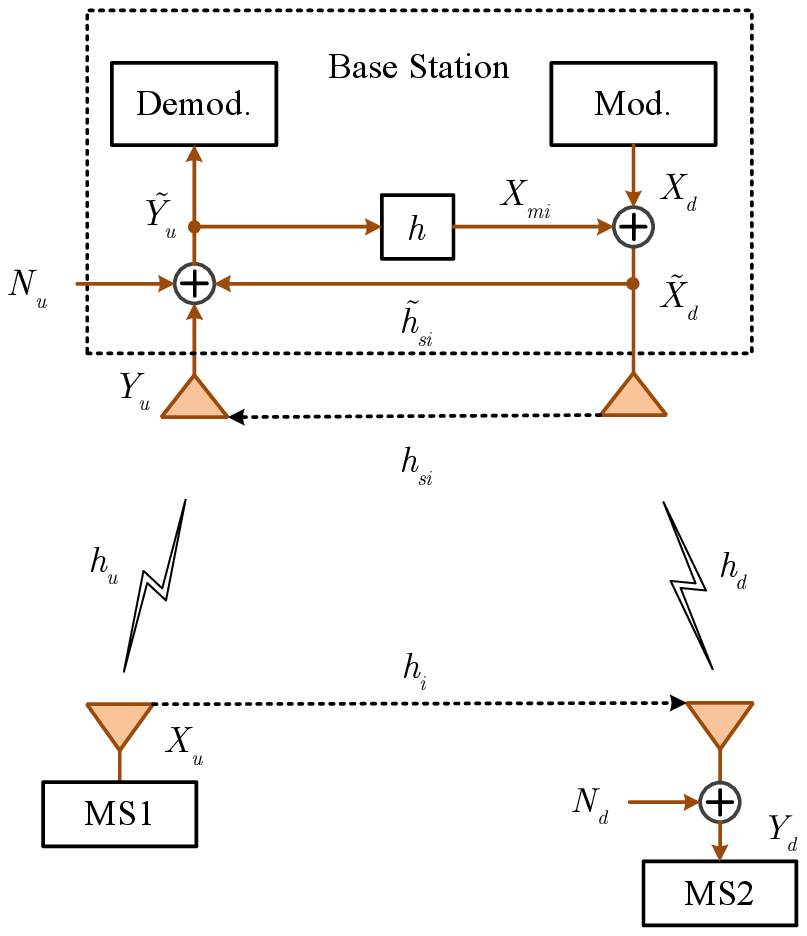}
\end{center}
\centering\caption{Three-node FD network.}
\label{f-Intern-model}
\end{figure}

In this section, we will introduce the system model to be used for the remainder of the paper.  As is shown in Fig. \ref{f-Intern-model}, the network is comprised of interference-free  uplink transmission and the downlink transmission interfered by the uplink user.

If the channels are fading, the signal model considered here is  frequency-flat and block-fading.  This implies that the system uses orthogonal frequency division multiplexing (OFDM) for broadband transmission over multipath channels. Thus, the signal model here represents a single narrowband subcarrier. The following equations give the signal relationship between BS and mobile users
\begin{align} \label{eqa:modleone}
Y_u^{} &= h_u{\left( \tau_u \right) }\ast X_u^{} +  \left( h_{si} \left( \tau_{si} \right) - \tilde{h}_{si}\left( \tilde{\tau}_{si} \right)  \right) \ast \tilde{X}_d^{}+ N_u^{}, \notag \\
\tilde{X}_d &= X_d + Y_u * h{\left( \tau \right) } , \notag \\
Y_d^{}& = h_d{\left( \tau_d \right) } \ast \tilde{X}_d^{} + h_i{\left( \tau_i \right) } \ast X_u^{} + N_d^{},
\end{align}
where $X_d$  and $X_d$  represent the uplink and the downlink signals respectively, and  $\Expect \left( \left| X_u \right| ^2 \right) \le P_u$  and   $\Expect \left( \left| X_d \right| ^2 \right) \le P_d$. Let $N_u$  and $N_d$  be independently and identically distributed (i.i.d.) complete white Gaussian noise with zero mean and variance of   $\sigma_n^2$.   $h_u$ denotes the direct link channel from the mobile user U1 to the BS, $h_d$  denotes the direct link channel from the BS to the mobile user U2, and  $h_i$ denotes the interference link channel from mobile user U1 to mobile user U2.  We assume that BS has the full channel station information of all other nodes. To simplify the notation, let  ${SNR}_{u}^{{}}=\frac{\left| h_{u}^{{}} \right|_{{}}^{2}P_{u}^{{}}}{\sigma _{n}^{2}}$,  ${SNR}_{d}^{{}}=\frac{\left| h_{d}^{{}} \right|_{{}}^{2}P_{d}^{{}}}{\sigma _{n}^{2}}$, and  ${INR}_{i}^{{}}=\frac{\left| h_{i}^{{}} \right|_{{}}^{2}P_{u}^{{}}}{\sigma _{n}^{2}}$.

The $\tilde {X}_d$ in  (\ref{eqa:modleone}) can be obtained by the following recursive equation
\begin{align} \label{eqa:modletwio}
\tilde{X}_d &= X_d +  \left(h_u{\left( \tau_u \right) }\ast X_u^{} +  \left( h_{si} \left( \tau_{si} \right) - \tilde{h}_{si}\left( \tilde{\tau}_{si} \right)  \right) \ast \tilde{X}_d^{}+ N_u^{} \right) * h{\left( \tau \right) },
\end{align}

For the wideband system, it is hard to solve the  (\ref{eqa:modletwio}) because of the recursion of the $\tilde {X}_d$. However, for the narrowband system, the channels can be modeled as single-path and the delay of the signal can be neglected and $\tilde {X}_d$ can be expressed as follows
\begin{align}
  \tilde{X}_{d}^{{}}  =\frac{ X_{d}^{{}}+hh_{u}^{{}}X_{u}^{{}}+hN_{u}^{{}}}{1 - \left( h_{si}  - \tilde{h}_{si}  \right) h}.
\end{align}
Thus, the received signals at the BS and the donwlink user can be written as

\begin{align}
Y_d &= \frac{{h_d^{}X_d^{}}}{{1 - \left( {h_{si}^{} - \tilde h_{si}^{}} \right)h}} + \left( {\frac{{hh_u^{}h_d^{}}}{{1 - \left( {h_{si}^{} - \tilde h_{si}^{}} \right)h}} + h_i^{}} \right)X_u^{} + \left( {\frac{{hh_d^{}N_u^{}}}{{1 - \left( {h_{si}^{} - \tilde h_{si}^{}} \right)h}} + N_d^{}} \right), \notag\\
Y_u &= \left( {h_u^{} + \frac{{hh_u^{}\left( {h_{si}^{} - \tilde h_{si}^{}} \right)}}{{1 - \left( {h_{si}^{} - \tilde h_{si}^{}} \right)h}}} \right)X_u^{} + \frac{{\left( {h_{si}^{} - \tilde h_{si}^{}} \right)X_d^{}}}{{1 - \left( {h_{si}^{} - \tilde h_{si}^{}} \right)h}} + \left( {\frac{{h\left( {h_{si}^{} - \tilde h_{si}^{}} \right)}}{{1 - \left( {h_{si}^{} - \tilde h_{si}^{}} \right)h}} + 1} \right)N_u^{}.
\end{align}

We assume the uplink signal  $X_u$, the downlink signal  $X_d$, and complex Gaussian noise  $N_u$ and $N_d$  are uncorrelated. Therefore, the signal-to-interference-plus-noise-ratios (SINRs) at the BS and the mobile user 2 can be expressed as
\[SINR_d{\left( h\right)} = \frac{{\left| {\frac{{h_d^{}}}{{1 - \left( {h_{si}^{} - \tilde h_{si}^{}} \right)h}}} \right|_{}^2P_d^{}}}{{\left| {\frac{{hh_u^{}h_d^{}}}{{1 - \left( {h_{si}^{} - \tilde h_{si}^{}} \right)h}} + h_i^{}} \right|_{}^2P_u^{} + \left( {\left| {\frac{{hh_d^{}}}{{1 - \left( {h_{si}^{} - \tilde h_{si}^{}} \right)h}}} \right|^2 + 1} \right)\sigma _n^2}}\]
and
\[SINR_u{\left( h\right)} = \frac{{\left| {h_u^{} + \frac{{hh_u^{}\left( {h_{si}^{} - \tilde h_{si}^{}} \right)}}{{1 - \left( {h_{si}^{} - \tilde h_{si}^{}} \right)h}}} \right|_{}^2P_u^{}}}{{\left| {\frac{{\left( {h_{si}^{} - \tilde h_{si}^{}} \right)}}{{1 - \left( {h_{si}^{} - \tilde h_{si}^{}} \right)h}}} \right|_{}^2P_d^{} + \left| {\frac{{h\left( {h_{si}^{} - \tilde h_{si}^{}} \right)}}{{1 - \left( {h_{si}^{} - \tilde h_{si}^{}} \right)h}} + 1} \right|_{}^2\sigma _n^2}}\]
respectively.

Therefore, the achievable rates of the downlink and the uplink can be written as
\[R_d^{} = \log _2^{}\left( {1 + SINR_d^{}\left( h \right)} \right)\]
and
\[R_u^{} = \log _2^{}\left( {1 + SINR_u^{}\left( h \right)} \right)\]
respectively.
Our optimization is to maximize the sum achievable rate as follows
\begin{align} \label{eqa:optimization}
R_{sum}^{} &= R_d^{} + R_u^{}\notag\\
 &= \log _2^{}\left( {1 + \frac{{\left| {\frac{{h_d^{}}}{{1 - \left( {h_{si}^{} - \tilde h_{si}^{}} \right)h}}} \right|_{}^2P_d^{}}}{{\left| {\frac{{hh_u^{}h_d^{}}}{{1 - \left( {h_{si}^{} - \tilde h_{si}^{}} \right)h}} + h_i^{}} \right|_{}^2P_u^{} + \left( {\left| {\frac{{hh_d^{}}}{{1 - \left( {h_{si}^{} - \tilde h_{si}^{}} \right)h}}} \right|^2 + 1} \right)\sigma _n^2}}} \right) + \notag\\
&\log _2^{}\left( {1 + \frac{{\left| {h_u^{} + \frac{{hh_u^{}\left( {h_{si}^{} - \tilde h_{si}^{}} \right)}}{{1 - \left( {h_{si}^{} - \tilde h_{si}^{}} \right)h}}} \right|_{}^2P_u^{}}}{{\left| {\frac{{\left( {h_{si}^{} - \tilde h_{si}^{}} \right)}}{{1 - \left( {h_{si}^{} - \tilde h_{si}^{}} \right)h}}} \right|_{}^2P_d^{} + \left| {\frac{{h\left( {h_{si}^{} - \tilde h_{si}^{}} \right)}}{{1 - \left( {h_{si}^{} - \tilde h_{si}^{}} \right)h}} + 1} \right|_{}^2\sigma _n^2}}} \right).
\end{align}

In the following section, we will propose a novel IUI suppression scheme for the narrowband system.For wideband system, we will introduce the scheme in Section IV based on the narrowband system.

Moreover, we introduce the Jensen inequality \cite{chung2001course}, which will  used throughout this paper, as follows
\begin{align}
\Expect \left[ \log_2 \left( 1 + X \right) \right] \le \log_2 \left( 1 + \Expect \left[  X \right] \right)
\end{align}
where $X$ is a nonnegative random variable.

 \section{IUI Suppression Scheme for narrowband system}

Since the joint optimization of the uplink transmit power $P_u$, the downlink transmit power $P_d$, and the IUI suppression coefficient $h$ under imperfect SI suppression at the BS cannot be derived by the explicit expression, we give the solution for different cases.

 \subsection{Optimization for Ideal Case}

We assume that  the SI at the BS is perfectly suppressed, the transmit power at the uplink user and the BS is fixed, and the BS has the perfect knowledge of wireless channels.

Since the BS knows the state information of all  channels  $h_u$, $h_d$,  and $h_i$, the IUI can be suppressed if the BS transmits the reversed version of the IUI signal as soon as it receives the uplink transmission signal. We assume the delay difference between the IUI signal and the reversed version of the IUI signal transmitted by the BS can be neglected\footnote{When the propagation distance difference between the IUI  and the IUI suppression signals is large, long OFDM frame should be adopted}.  Here we assume the SI is perfectly or almost mitigated, i.e., $h_{si} = \tilde{h}_{si}$. Then, the optimization objective becomes
 \begin{align}
R_{sum}^{} &=  \log _2^{}\left( 1 +  {\frac{{\left| {h_d^{}} \right|_{}^2P_d^{}}}{{\left| {hh_u^{}h_d^{} + h_i^{}} \right|_{}^2P_u^{} + \left( {\left| {hh_d^{}} \right| + 1} \right)\sigma _n^2}}}\right) + \notag\\
&\log _2^{}\left(1 +\frac{{\left| {h_u^{}} \right|_{}^2P_u^{}}}{{\sigma _n^2}} \right).
\end{align}

Obviously, the optimization can be equivalent to maximize the $SINR_d$ as follows
\[{SINR_d=}\frac{\left| h_{d}^{{}} \right|_{{}}^{2}P_{d}^{{}}}{\left| hh_{d}^{{}}h_{u}^{{}}+h_{i}^{{}} \right|_{{}}^{2}P_{u}^{{}}+\left| h_{d}^{{}}h \right|_{{}}^{2}\sigma _{n}^{2}+\sigma _{n}^{2}}\]

Let interference-plus-noise-power (INP) be  $INP=\left| hh_{d}^{{}}h_{u}^{{}}+h_{i}^{{}} \right|_{{}}^{2}P_{u}^{{}}+\left| h_{d}^{{}}h \right|_{{}}^{2}\sigma _{n}^{2}+\sigma _{n}^{2}$. The problem to maximize the $SINR$   is equivalent to minimize   $INP$ as follows
\[\underset{h\in \mathbb{C}}{\mathop{\min }}\,{INP}=\underset{h\in \mathbb{C}}{\mathop{\min }}\,\left( \left| hh_{d}^{{}}h_{u}^{{}}+h_{i}^{{}} \right|_{{}}^{2}P_{u}^{{}}+\left| hh_{d}^{{}} \right|_{{}}^{2}\sigma _{n}^{2}+\sigma _{n}^{2} \right)\]

Differentiate  $INP$ once yields
\[\frac{\partial {INP}}{\partial h}=2P_{u}^{{}}\left( h\left| h_{d}^{{}}h_{u}^{{}} \right|_{{}}^{2}+h_{i}^{{}}h_{d}^{*}h_{u}^{*} \right)+2h\sigma _{n}^{2}\left| h_{d}^{{}} \right|_{{}}^{2}\]

Let the differentiation be zero, we get the optimum $h$  as follows
\begin{align}
h_{{opt}}^{{}}=-\frac{h_{i}^{{}}h_{d}^{*}h_{u}^{*}P_{u}^{{}}}{\left| h_{d}^{{}}h_{u}^{{}} \right|_{{}}^{2}P_{u}^{{}}+\left| h_{d}^{{}} \right|_{{}}^{2}\sigma _{n}^{2}}.
\end{align}

Therefore, the maximization   output at the mobile user U2 is
\begin{align}
{SINR}_{{opt}}^{{}} &=\frac{\left| h_{d}^{{}} \right|_{{}}^{2}P_{d}^{{}}}{\left| \frac{h_{i}^{{}}\sigma _{n}^{2}}{\left| h_{u}^{{}} \right|_{{}}^{2}P_{u}^{{}}+\sigma _{n}^{2}} \right|_{{}}^{2}P_{u}^{{}}+\left| \frac{h_{i}^{{}}h_{u}^{*}P_{u}^{{}}}{\left| h_{u}^{{}} \right|_{{}}^{2}P_{u}^{{}}+\sigma _{n}^{2}} \right|_{{}}^{2}\sigma _{n}^{2}+\sigma _{n}^{2}} \notag\\
&=\frac{\left| h_{d}^{{}} \right|_{{}}^{2}P_{d}^{{}}}{\frac{\left| h_{i}^{{}} \right|_{{}}^{2}P_{u}^{{}}\sigma _{n}^{2}}{\left| h_{u}^{{}} \right|_{{}}^{2}P_{u}^{{}}+\sigma _{n}^{2}}+\sigma _{n}^{2}}
\end{align}

In above analysis, we assume the transmit power of BS is not limited. Thus, there exists enough power to generate the reversed version of the IUI signal.

\emph{Remark 1:}
 We can see that the residual IUI denoted as $INP$  at the mobile user U2  avoids from the influence of the downlink channel power $\left|h_d\right|^2$ . In addition,  $SINR_{opt}$  is increased with increase of    the downlink channel power  $\left|h_d\right|^2$.

\emph{Remark 2:}
For fixed  $\left|h_d\right|^2$, it is obvious that the ratio of the interference channel and the uplink channel power, \ie, $\left|h_d/h_u\right|^2$, decides the  $SINR_{opt}$. Specially, when  $P_d = P_u \gg \sigma_n^2$ and  $\left|h_d\right|^2 = 1$, if  $\left|h_i\right|^2 = \left|h_u\right|^2 = 1$ ( $\left|h_d/h_u\right|^2 = 1$ ) , it means that  $SINR_{opt}$  degrades about 3dB compared to the IUI-free case. On the other hand, this rate  is in one-bit rate scope of the ideal case when there is no IUI.

\emph{Remark 3:}
If  $\left|h_i\right|^2 \ll \left|h_u\right|^2$, it means that  $SINR_{opt}$  approximates to performance of IUI-free situation. Even when  $\left|h_i\right|^2 \gg \left|h_u\right|^2$, the IUI can be suppressed,   as long as the channel power of the uplink $\left|h_u\right|^2$ is not zero.  Specially, when the uplink signal power received by the BS $\left|h_u\right|^2P_u $ is equal to noise floor $\sigma _{n}^{2}$, the IUI can be attenuated half. The reason is that the rate of  noise floor increase at the downlink user is slower than the rate of the IUI reduction when applying the proposed IUI suppression method. For example, if the IUI is $10x$ and $\Expect \left(x \right) =1$, then the power of the IUI is $100$. When the received signal at the BS is $x$ and the uncorrelated noise floor  is $n$ where $\Expect \left(n \right) =1$, subtracting the IUI $10x$ by $5x + 5n$, which  is five times of the signal received at the BS,  yields $5x + 5n$ and the residual IUI power is attenuated half to $50$.


 \subsection{Practical Considerations}
 In the following, we consider some practical restrictions, such as  the imperfect SI suppression of the BS , the delay difference between the IUI and the IUI suppression signals, the limited total transmit power of the BS, the power control of the uplink and downlink, and imperfect channel information, and analyze the their impact on the performance of the proposed IUI suppression scheme.

 \subsubsection{Imperfect SI suppression of the BS} If we take the residual SI signal into account, the  maximum optimization of the sum rate in (\ref{eqa:optimization}) is hard to give the close-form expression.  Here we just discuss the trending of sum rate. We assume the optimization of the adjustable coefficient is $h_{opt}$, then the sum rate $R_{sum}$ is
 \begin{align}
R_{sum}^{} &= R_d^{} + R_u^{}\notag\\
 &= \log _2^{}\left( {1 + \frac{{\left| {\frac{{h_d^{}}}{{1 - \left( {h_{si}^{} - \tilde h_{si}^{}} \right)h_{opt}}}} \right|_{}^2P_d^{}}}{{\left| {\frac{{hh_u^{}h_d^{}}}{{1 - \left( {h_{si}^{} - \tilde h_{si}^{}} \right)h_{opt}}} + h_i^{}} \right|_{}^2P_u^{} + \left( {\left| {\frac{{hh_d^{}}}{{1 - \left( {h_{si}^{} - \tilde h_{si}^{}} \right)h_{opt}}}} \right|^2 + 1} \right)\sigma _n^2}}} \right) + \notag\\
&\log _2^{}\left( {1 + \frac{{\left| {h_u^{} + \frac{{hh_u^{}\left( {h_{si}^{} - \tilde h_{si}^{}} \right)}}{{1 - \left( {h_{si}^{} - \tilde h_{si}^{}} \right)h_{opt}}}} \right|_{}^2P_u^{}}}{{\left| {\frac{{\left( {h_{si}^{} - \tilde h_{si}^{}} \right)}}{{1 - \left( {h_{si}^{} - \tilde h_{si}^{}} \right)h_{opt}}}} \right|_{}^2P_d^{} + \left| {\frac{{h_{opt} \left( {h_{si}^{} - \tilde h_{si}^{}} \right)}}{{1 - \left( {h_{si}^{} - \tilde h_{si}^{}} \right)h_{opt}}} + 1} \right|_{}^2\sigma _n^2}}} \right).
\end{align}

\emph{Remark 1:} It is easy to know that the sum rate $R_{sum}$ is increased with the SI suppression capability, i.e, $R_{sum} \propto \frac{1}{\left|\left( {h_{si}^{} - \tilde h_{si}^{}} \right)\right|^2}$.
Specially, when no SI suppression operation is carrier out, i.e., $\tilde h_{si} =0$, then the uplink rate $R_u \approx 0$.
 On the other hand, the SNR of the received uplink signal at the BS is too small.
 In this case, the proposed scheme will be ineffective.


 \subsubsection{Delay difference between the IUI and the IUI suppression signals}
Obviously, the delay difference  $\tau_{i,d} = \tau_i - \tilde{\tau}_i$ between the IUI signal delay $\tau_i$  and the IUI suppressio signal delay $\tilde{\tau}_i$   affects the IUI suppression effects. Assuming the modulation signals is $a \left( n \right)$ and $\Expect \left[ a \left( n \right)a^\ast \left( n \right)\right]=1$, then  the uplink signals $ X_u^{}$ at the baseband can be expressed as
\begin{align}
X_u = \sum\limits_{n=-\infty}^{\infty} a \left( n \right) p_T \left( t - nT \right),
\end{align}
where $T$ denotes the symbol duration and $p_T\left( t \right)$ denotes the signal shaping filter. Here we adopt raised cosine roll-off filter as follows
\begin{align}
p_T\left( t \right) =\left\{ \begin{array}{l}
\frac{\pi }{{4T}}{\rm{sinc}}\left( {\frac{1}{{2\beta }}} \right),\;\;\;\;\;\;\;\;\;\;\;\;t =  \pm \frac{T}{{2\beta }}\\
\frac{1}{T}{\rm{sinc}}\left( {\frac{t}{T}} \right)\frac{{\cos \left( {\frac{{\pi \beta t}}{T}} \right)}}{{1 - \left( {\frac{{2\beta t}}{T}} \right)_{}^2}},\;\;{\rm{otherwise}}\;
\end{array} \right.
\end{align}
where $  0 \le \beta \le 1$ denotes the \emph{roll-off factor}.

For simplify, we only consider the IUI suppression in one symbol.
Thus, the IUI sinal at the downlink user is
\begin{align}
Y_{IUI}& = h_i\delta\left( t - \tau_i \right) \otimes X_u \notag\\
 &=   h_i a \left( n \right) p_T \left( t -\tau_i - nT \right),
\end{align}
where $\delta\left(\cdot\right)$ denotes the Dirac Delta function.

Similarly, the IUI suppression at the downlink user is
\begin{align}
Y_{IUIS}& = \tilde{h}_i\delta\left( t - \tilde{\tau}_i \right) \otimes X_u \notag\\
 &=   \tilde{h}_i a \left( n \right) p_T \left( t -\tilde{\tau}_i - nT \right),
\end{align}

Neglecting the additive noise and assuming $\tilde{h}_i = -{h}_i$ and $\left|\tilde{h}_i\right|  = 1$,  the residual IUI signal after IUI suppression at the downlink user approximately is
\begin{align}
\tilde{Y}_{IUI} & = Y_{IUI} - Y_{IUIS} \notag\\
          &= h_i  a \left( n \right) \left( p_T \left( t -\tau_i - nT \right) - p_T \left( t -\tilde{\tau}_i - nT \right)\right)
\end{align}

The  IUI suppressio ratio can be expressed as

\begin{align}
r\left(\tau_{i,d} \right) &= \frac{\Expect\left[ \tilde{Y}_{IUI}\tilde{Y}_{IUI}^\ast\right]}{\Expect\left[ Y_{IUI}Y_{IUI}^\ast\right]} \notag\\
                         &= 2\frac {R_p \left( 0 \right) - R_p \left( \tau_{i,d}  \right)}{ R_p \left( 0 \right)}
\end{align}
where $R_p \left( \tau \right)$ denotes the  auto-correlation function of raised cosine function $p_T\left( t \right)$  as follows
\[R\left( \tau  \right) = T\left( {{\rm{sinc}}\left( {\frac{\tau }{T}} \right)\frac{{\cos \left( {\frac{{\pi \beta \tau }}{T}} \right)}}{{1 - \left( {\frac{{2\beta \tau }}{T}} \right)_{}^2}} - \frac{\beta }{4}{\rm{sinc}}\left( {\frac{{\beta \tau }}{T}} \right)\frac{{\cos \left( {\frac{{\pi \tau }}{T}} \right)}}{{1 - \left( {\frac{{\beta \tau }}{T}} \right)_{}^2}}} \right)\]

\begin{figure}
\begin{center}
\includegraphics[width=0.6\textwidth]{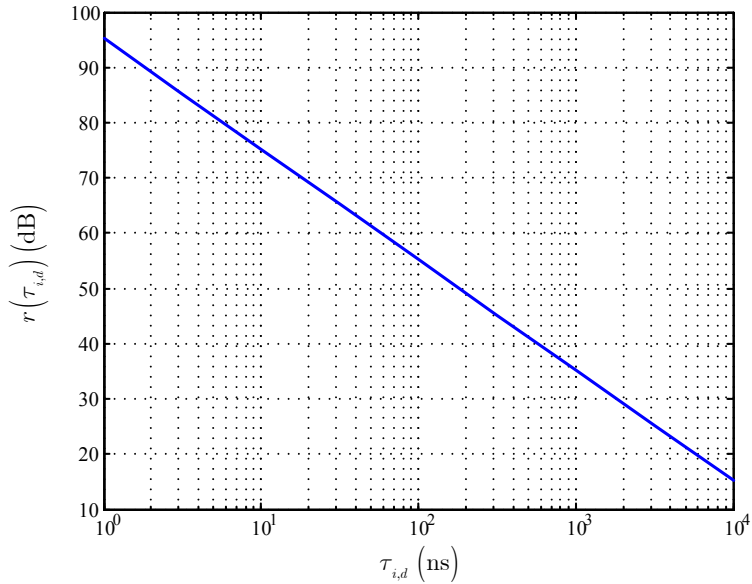}
\end{center}
\centering\caption{The IUI suppression ratio $r \left( \tau_{i,d}\right)$ vs. the delay difference $\tau_{i,d}$. $T = 100 \mu s$ and $\beta = 0.22$.}
\label{f-Delaydifference}
\end{figure}

In the following, we investigate the impact of the delay difference $\tau_{i,d}$ on the IUI suppression. Here, we set $T = 100 \mu s$ and $\beta = 0.22$.
Fig. \ref{f-Delaydifference} shows the the IUI suppression ratio $r \left( \tau_{i,d}\right)$ versus the delay difference $\tau_{i,d}$. When the delay difference $\tau_{i,d}=2 \mu s$, \ie., the $\frac{\tau_{i,d}}{T} = \frac{1}{50}$, the IUI suppression ratio is about $30$dB.
That is to say,  long enough symbol duration is needed to maintain the IUI suppression of the proposed scheme when the delay difference is fixed.

 \subsubsection{Power control of the uplink and downlink}

If power control of the uplink and downlink is considered, the performance of the proposed INI suppression scheme may be further improved. In the following, we discuss it for two cases.

\textbf{Case 1:} Perfect SI mitigation at the BS.

For this case, the sum achievable rate $R_{sum}^{}$ can be expressed as
 \begin{align}
R_{sum}^{} &=  \log _2^{}\left( 1 +  {\frac{\left| h_{d}^{{}} \right|_{{}}^{2}P_{d}^{{}}}{\frac{\left| h_{i}^{{}} \right|_{{}}^{2}P_{u}^{{}}\sigma _{n}^{2}}{\left| h_{u}^{{}} \right|_{{}}^{2}P_{u}^{{}}+\sigma _{n}^{2}}+\sigma _{n}^{2}}}\right) +\log _2^{}\left(1 +\frac{{\left| {h_u^{}} \right|_{}^2P_u^{}}}{{\sigma _n^2}} \right).
\end{align}

\emph{Remark 1:} It is obviously that the sum achievable rate  $R_{sum}^{}$ is increased with the downlink transmit power $P_d$. Thus, the downlink transmit power $P_d$ should be set to the largest.

\emph{Remark 2:} The first term of the $R_{sum}^{}$ is decreased with the uplink transmit power $P_u$ and the second term is increased with  $P_u$. Thus, the uplink transmit power $P_u$ may exist a optimum transmit power between the $0$ and the maximum uplink transmit power. Specially, when the received SNR of the uplink is much greater than one, i.e., $SNR ={\left| h_{u}^{{}} \right|_{{}}^{2}P_{u}^{{}}/\sigma _{n}^{2}} \gg 1$, we can get $R_{sum}^{} \approx  \log _2^{}\left( 1 +  {\frac{\left| h_{d}^{{}} \right|_{{}}^{2} {\left| h_{u}^{{}} \right|_{{}}^{2}} P_{d}^{{}}}{{\left| h_{i}^{{}} \right|_{{}}^{2}\sigma _{n}^{2}}+{\left| h_{u}^{{}} \right|_{{}}^{2}}\sigma _{n}^{2}}}\right) +\log _2^{}\left(1 +\frac{{\left| {h_u^{}} \right|_{}^2P_u^{}}}{{\sigma _n^2}} \right)$. Obviously, the sum achievable rate $R_{sum}^{}$ is increased with the uplink transmit power $P_u$. In this situation, the uplink transmit power $P_u$ should be set to the largest.

\textbf{Case 2:} Imperfect SI mitigation at the BS.

 Since the INI suppression signal is retransmitted for this case, it is hard to give the explicit expression.

   Up to data, many researchers have devoted to INI suppression using the power control, please refer to  \cite{zhang2015investigation,feng2015joint,imari2016theoretical,li2016binary,mairton2016distributed} and the referees therein.

 Specially, a simple power control scheme is proposed in \cite{li2016binary}.
 In this literature, the SI and the INI are both taken into account for power control optimization in single-cell and multi-cell FD networks.
 They show that the binary feature in sum rate-optimized power control solution holds, even when applying the INI suppression techniques referred by the method used in the multi-access channel.
 Thus, the proposed INI suppression scheme can be companied with this power control scheme.
 Since this study is too complicated, we leave it for future work.

 \subsubsection{Imperfect channel information}

Usually, we have perfect knowledge of the uplink channel $h_u$  and imperfect channel information of the downlink and IUI channels because of limited feedback bandwidth. Assuming that we the perfect amplitudes of the downlink and IUI channels and the their phases obeys $\left[ 0, 2 \pi \right)$ uniform distribution, i.e., $h_d = \left|h_d\right| e^{j\theta_d}$ and $h_i = \left|h_i\right| e^{j\theta_i}$, $\theta_d \sim U\left[-\pi, \pi \right)$, $\theta_i \sim U\left[-\pi, \pi \right)$, where $U\left[a, b\right)$ represents the uniform distribution in interval $\left(a, b \right)$.

Let $h = \left|h\right| e^{j\theta}$ and $h_u = \left|h\right| e^{j\theta_u}$, where $\theta$ and $\theta_u$ are specific phase values. Thus, the average INP becomes
\begin{align}
INP &= \Expect \left[\left| hh_{d}^{{}}h_{u}^{{}}+h_{i}^{{}} \right|_{{}}^{2}P_{u}^{{}}+\left| h_{d}^{{}}h \right|_{{}}^{2}\sigma _{n}^{2}+\sigma _{n}^{2} \right] \notag\\
    &= 2\left| {h_i^{}} \right|\left| h \right|\left| {h_d^{}} \right|\left| {h_u^{}} \right| \Expect\left[ \cos \left( {\theta _i^{}  - \theta _d^{}-\theta  - \theta _u^{}} \right)\right]P_u^{} + \left| h \right|_{}^2\left| {h_d^{}} \right|_{}^2\left| {h_u^{}} \right|_{}^2P_u^{} \notag\\
     &+\left| {h_i^{}} \right|_{}^2P_u^{}+\left| h_{d}^{{}}\right|^2 \left|h \right|_{{}}^{2}\sigma _{n}^{2}+\sigma _{n}^{2}
\end{align}

Let $\gamma = \alpha  -\beta$ and $\alpha\sim U\left[-\pi, \pi \right) $ and $\beta \sim U\left[-\pi, \pi \right)$ are independent. The distribution function of the $\gamma$ can be calculated as follows:

When $\gamma \ge 0$,
\begin{align}
  F_\gamma ^{}\left( \gamma  \right) &= \iint\limits_{\alpha  - \beta  < \gamma } {\frac{1}{{4\pi _{}^2}}}d\alpha d\beta  \notag \\
   &= 1 - \iint\limits_{\alpha  - \beta \geq   \gamma } {\frac{1}{{4\pi _{}^2}}}d\alpha d\beta  \notag \\
   &= 1 - \int_{\gamma  - \pi }^\pi  {\int_{ - \pi }^{\alpha  - \gamma } {d\beta d\alpha } }  \notag \\
   &= \frac{1}{2} + \frac{\gamma }{{2\pi }} - \frac{{\gamma _{}^2}}{{8\pi _{}^2}}
\end{align}

When $\gamma < 0$,

\begin{align}
      F_\gamma ^{}\left( \gamma  \right) &= \iint\limits_{\alpha  - \beta  < \gamma } {\frac{1}{{4\pi _{}^2}}}d\alpha d\beta  \notag \\
   &= \frac{1}{{4\pi _{}^2}}\int_{ - \pi }^{\pi  + \gamma } {\int_{\alpha  - \gamma }^\pi  {d\beta } } d\alpha  \notag \\
   &= \frac{1}{2} + \frac{\gamma }{{2\pi }} + \frac{{\gamma _{}^2}}{{8\pi _{}^2}}
\end{align}

The probability distribution function (PDF) of $\gamma$ is
\begin{align}
f_\gamma ^{}\left( \gamma  \right) = \left\{ \begin{gathered}
  \frac{1}{{2\pi }} - \frac{\gamma }{{4\pi _{}^2}},\gamma  \geq 0 \hfill \\
  \frac{1}{{2\pi }} + \frac{\gamma }{{4\pi _{}^2}},\gamma  < 0 \hfill \\
\end{gathered}  \right.
\end{align}

Thus, the expectation of the ${\cos \left( {\gamma  - \theta_f } \right)}$, where $\theta_f$ is a constant, is
\begin{align}
\Expect \left[ {\cos \left( {\gamma  - \theta_f } \right)} \right] = \int_{ - 2\pi }^{2\pi } {\cos \left( {\gamma  - \theta_f } \right)f_\gamma ^{}\left( \gamma  \right)d\gamma }  = 0
\end{align}

Therefore, the INP becomes
\begin{align}
INP =\left| {h_i^{}} \right|_{}^2P_u^{}+\left| h_{d}^{{}}\right|^2 \left|h \right|_{{}}^{2}\sigma _{n}^{2}+\sigma _{n}^{2}
\end{align}
It is obviously that INP reaches the minimum when the amplitude of the weighted coefficient $h$ is zero, i.e., $\left|h\right| =0$.  Notice that  this conclusion holds on when any of the phase of the  the downlink and interference channels is uniformly random.

\emph{Remark 1:} The proposed INI suppression scheme fails in the situation when the phase information of the cam not be obtained. On the other hand, partial or precise phase information of the downlink and interference channels at the BS is necessary.

 \subsubsection{Limited total transmit power}
  In practice, the total transmit power is limited. In the following, we total available power is $P$. For simplify, the uplink and downlink power is the same.

   We assume that the uplink, the downlink, and the interference channels are Rayleigh fading. We assume that the transceiver transmits signals in uniform power. We also assume that $h_i \in \mathcal{CN} \left( 0, 1 \right)$. $h_d \in \mathcal{CN} \left( 0, 1 \right)$ and $h_i \in \mathcal{CN} \left( 0, 1 \right)$.

   To evaluate the performance of proposed IUI suppression scheme, four scenarios are considered as follows.
   Case one, the networks works in HD mode.
   Case two, the networks works in ideal FD mode without IUI.
   Case three, the networks works in practical FD mode with IUI but no any IUI suppression technique
is adopted.
  Case four, the networks works in practical FD mode with IUI suppressed by the proposed IUI scheme.

   In the following, we first give the sum achievable rate of the four cases and then energy efficiency.

\textbf{a):} Sum achievable rate

For case one, the network works in HD mode. We assume the uplink ratio is  $\mu ,0\le \mu \le 1$, then the downlink ratio is   $\left( 1-\mu  \right)$.  The sum  achievable rate is
\begin{align} \label{eqa:rayleirate1}
R_{sum}^{HD}=& \Expect \left[ \mu \log _{2}^{{}}\left( 1+\frac{\left| h_{u}^{{}} \right|_{{}}^{2}P_{u}^{{}}}{\sigma _{n}^{2}} \right)+ \right. \notag\\
&\left.\left( 1-\mu  \right)\log _{2}^{{}}\left( 1+\frac{\left| h_{d}^{{}} \right|_{{}}^{2}P_{d}^{{}}}{\sigma _{n}^{2}} \right) \right] \notag\\
 & \le \mu \log _{2}^{{}}\left( 1+\frac{P_u}{\sigma _{n}^{2}} \right)+\left( 1-\mu  \right)\log _{2}^{{}}\left( 1+\frac{P_d}{\sigma _{n}^{2}} \right).
\end{align}

For case two, the network works in FD mode while no IUI exists. In practice, this case occurs when the downlink user is hidden from the uplink user. The sum achievable rate is
\begin{align} \label{eqa:rayleirate2}
R_{sum}^{FD,1}=&\Expect \left[ \log _{2}^{{}}\left( 1+\frac{\left| h_{u}^{{}} \right|_{{}}^{2}P_{u}^{{}}}{\sigma _{n}^{2}} \right)+ \right. \notag\\
&\left.\log _{2}^{{}}\left( 1+\frac{\left| h_{d}^{{}} \right|_{{}}^{2}P_{d}^{{}}}{\sigma _{n}^{2}} \right) \right] \notag\\
 & \le \log _{2}^{{}}\left( 1+\frac{P_u}{\sigma _{n}^{2}} \right)+\log _{2}^{{}}\left( 1+\frac{P_d}{\sigma _{n}^{2}} \right).
\end{align}

For case three, the sum achievable rate is

\begin{align}
   R_{sum}^{FD,{2}}=& \Expect\left[ \log _{2}^{{}}\left( 1+\frac{\left| h_{u}^{{}} \right|_{{}}^{2}P_{u}^{{}}}{\sigma _{n}^{2}} \right)+ \right. \notag\\
   &\left.\log _{2}^{{}}\left( 1+\frac{\left| h_{d}^{{}} \right|_{{}}^{2}P_{d}^{{}}}{\left| h_{i}^{{}} \right|_{{}}^{2}P_{u}^{{}}+\sigma _{n}^{2}} \right) \right] \notag\\
 & \le \log _{2}^{{}}\left( 1+\frac{P_{u}^{{}}}{\sigma _{n}^{2}} \right)+\notag\\
 &\log_{2}^{{}}\left( 1-\frac{P_{d}^{{}}}{P_{u}^{{}}}\exp \left( \frac{\sigma _{n}^{2}}{P_{u}^{{}}} \right)\Ei\left( -\frac{\sigma _{n}^{2}}{P_{u}^{{}}} \right) \right).
\end{align}

For  case four, sometimes in practice, if the uplink channel power $\left| h_u \right|^2$ is small, it needs to control the usage of the proposed IUI suppression method. Thus, we introduce the threshold coefficient of the uplink channel power $\beta, \beta \ge 0$ which decides whether the BS attempts to or not suppress the IUI\footnote{When the threshold coefficient of the uplink channel power $\beta =0$, the results are in theory. In the following, we set $\beta =1$. Moreover,  the simulation results are similar to the  situation when  $\beta =0$ in our parameter set.}. If $ \left| h_u \right|^2 \ge \beta \sigma_n^2 / P_u$, the BS decides to transmit the extra signal to suppress the IUI. In other situation, the BS keeps silent.   The sum achievable rate of three nodes network can be expressed as

\begin{align} \label{equ:rayleigh}
   R_{{sum}}^{{F}D,{3}}=& \Expect\left[ \log _{2}^{{}}\left( 1+\frac{\left| h_{u}^{{}} \right|_{{}}^{2}P_{u}^{{}}}{\sigma _{n}^{2}} \right)+ \right.  \notag\\
   &\log _{2}^{{}}\left( 1+\left. \frac{\left| h_{d}^{{}} \right|_{{}}^{2}P_{d}^{{}}}{\frac{\left| h_{i}^{{}} \right|_{{}}^{2}P_{u}^{{}}\sigma _{n}^{2}}{\left| h_{u}^{{}} \right|_{{}}^{2}P_{u}^{{}}+\sigma _{n}^{2}}+\sigma _{n}^{2}} \right| \left|  h_{u}^{{}} \right|_{{}}^{2}>\frac{\beta \sigma _{n}^{2}}{P_{u}^{{}}}  \right) + \notag\\
    & \left.\log _{2}^{{}}\left( 1+ \left.\frac{\left| h_{d}^{{}} \right|_{{}}^{2}P_{d}^{{}}}{\left| h_{i}^{{}} \right|_{{}}^{2}P_{u}^{{}}+\sigma _{n}^{2}}\right| \left| h_{u}^{{}} \right|_{{}}^{2}\le \frac{\beta \sigma _{n}^{2}}{P_{u}^{{}}} \right) \right] \notag \\
 & \le \log _{2}^{{}}\left( 1+\frac{P_{u}^{{}}}{\sigma _{n}^{2}} \right)+\notag \\
 &\log _{2}^{{}}\left( 1+\left. \Expect\left[ \frac{\left| h_{d}^{{}} \right|_{{}}^{2}P_{d}^{{}}}{\frac{\left| h_{i}^{{}} \right|_{{}}^{2}P_{u}^{{}}\sigma _{n}^{2}}{\left| h_{u}^{{}} \right|_{{}}^{2}P_{u}^{{}}+\sigma _{n}^{2}}+\sigma _{n}^{2}} \right|\left|  h_{u}^{{}} \right|_{{}}^{2}>\frac{\beta \sigma _{n}^{2}}{P_{u}^{{}}}\right]\right)+ \notag\\
 &\log _{2}^{{}}\left( 1+\left. \Expect\left[ \frac{\left| h_{d}^{{}} \right|_{{}}^{2}P_{d}^{{}}}{\left| h_{i}^{{}} \right|_{{}}^{2}P_{u}^{{}}+\sigma _{n}^{2}} \right|\left|  h_{u}^{{}} \right|_{{}}^{2}\le\frac{\beta \sigma _{n}^{2}}{P_{u}^{{}}}\right] \right)
\end{align}

\begin{figure}
\begin{center}
\includegraphics[width=0.6\textwidth]{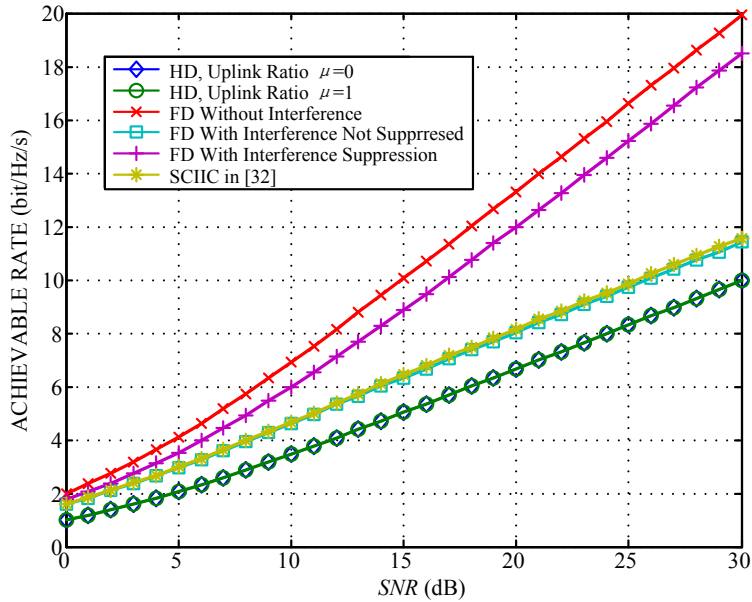}
\end{center}
\centering\caption{Sum achievable rate vs. $SNR$. $SNR_u = SNR_d = INR_i = SNR$. The threshold coefficient of the uplink channel power $\beta = 1$.}
\label{f-FadingRayleighRate}
\end{figure}

Fig. \ref{f-FadingRayleighRate} shows the sum achievable rate versus SNR  for five cases besides the SCIIC scheme in \cite{bi2016superposition}.
The gap between the ideal FD mode without interference and the FD mode with IUI suppression by the proposed scheme in this paper almost keep constant from  $0$ dB to $30$ dB SNR, where $SNR_u = SNR_d = INR_i = SNR$.
In addition, the proposed scheme performs better than SCIIC.
However, the difference between the FD mode with IUI suppression by the proposed scheme and  the FD mode with interference not suppressed increases with the increase of the SNR.
Moreover, even the FD mode with interference not suppressed performs better than the HD mode.
This result verifies that the FD operation can obtain benefit even the IUI is considered as an additive Gaussian noise at the U2 mobile receiver.

\textbf{b):} Energy efficiency

In the following, we assume   $P_{d}^{{}}+P_{u}^{{}}+P_{j}^{{}}=P$  where  $P_{j}^{{}}$  is the average power used for suppressing the IUI. For simplify, we assume  $P_{u}^{{}}=P_{d}^{{}}$.

For case one, it easy to know that $P_{u}^{{}}=P_{d}^{{}}={P}/{2}\;$, thus the energy efficiency is
\begin{align} \label{eqa:rayleighengy1}
E_{HD}^{{}}&=\frac{R_{sum}^{HD}}{P} \notag\\
&=\frac{\mu \log _{2}^{{}}\left( 1+\frac{P}{2\sigma _{n}^{2}} \right)+\left( 1-\mu  \right)\log _{2}^{{}}\left( 1+\frac{P}{2\sigma _{n}^{2}} \right)}{P}.
\end{align}

For case two,  also, $P_{u}^{{}}=P_{d}^{{}}={P}/{2}\;$, thus the energy efficiency is
\begin{align} \label{eqa:rayleighengy2}
E_{FD}^{1}&=\frac{R_{sum}^{FD,1}}{P} \notag\\
&=\frac{\log _{2}^{{}}\left( 1+\frac{P}{2\sigma _{n}^{2}} \right)+\log _{2}^{{}}\left( 1+\frac{P}{2\sigma _{n}^{2}} \right)}{P}
\end{align}

For case three,  also, $P_{j}^{{}}=0$  and $P_{d}^{{}}=P_{u}^{{}}={P}/{2}\;$. The energy efficiency is
\begin{align}
E_{FD}^{{2}}&=\frac{R_{sum}^{FD,{2}}}{P}\notag\\
&=\frac{\log _{2}^{{}}\left( 1+\frac{1}{\check{P} } \right)+\log _{2}^{{}}\left( 1-\exp \left( \check{P}  \right)\Ei\left( -\check{P}\right) \right)}{P}\notag
\end{align}
where $\check{P} = \frac{2\sigma _{n}^{2}}{P}$.

For case four, the energy efficiency is
\begin{align}
    E_{FD}^{3}=\frac{R_{{sum}}^{{F}D,{3}}}{P}
\end{align}
where
\begin{align} \label{eqa:rayleienery1}
   R_{{sum}}^{{F}D,{3}}=&\Expect\left[ \left. \log _{2}^{{}} \left( 1+\frac{\left| h_{u}^{{}} \right|_{{}}^{2}P_{u}^{{}}}{\sigma _{n}^{2}} \right) \right| \left| h_{u}^{{}} \right|_{{}}^{2}<\frac{\beta \sigma _{n}^{2}}{P_{u}^{{}}} \,  \mathrm{or} \, \left| h_{d}^{{}} \right|_{{}}^{2}<T_{d}^{{}} \right. + \notag\\
   & \left.\log_2 \left( 1+  \frac{\left| h_{d}^{{}} \right|_{{}}^{2}P_{d}^{{}}}{\left| h_{i}^{{}} \right|_{{}}^{2}P_{u}^{{}}+\sigma _{n}^{2}}  \right)\right| \left| h_{u}^{{}} \right|_{{}}^{2}<\frac{\beta \sigma _{n}^{2}}{P_{u}^{{}}} \,  \mathrm{or} \, \left| h_{d}^{{}} \right|_{{}}^{2}<T_{d}^{{}}  + \notag\\
 & \left.   \log _{2}^{{}} \left( 1+\frac{\left| h_{u}^{{}} \right|_{{}}^{2}P_{u}^{{}}}{\sigma _{n}^{2}} \right)\right|\left| h_{u}^{{}} \right|_{{}}^{2}\ge \frac{\beta \sigma _{n}^{2}}{P_{u}^{{}}},\left| h_{d}^{{}} \right|_{{}}^{2}\ge T_{d}^{{}}+ \notag\\
 &\left. \left.\log _{2}^{{}}\left( 1+ \frac{\left| h_{d}^{{}} \right|_{{}}^{2}P_{d}^{{}}}{\frac{\left| h_{i}^{{}} \right|_{{}}^{2}P_{u}^{{}}\sigma _{n}^{2}}{\left| h_{u}^{{}} \right|_{{}}^{2}P_{u}^{{}}+\sigma _{n}^{2}}+\sigma _{n}^{2}} \right) \right|\left| h_{u}^{{}} \right|_{{}}^{2}\ge \frac{\beta \sigma _{n}^{2}}{P_{u}^{{}}},\left| h_{d}^{{}} \right|_{{}}^{2}\ge T_{d}^{{}}  \right]
\end{align}

For case four,  the transmitted power  $P_j$  can be obtained by
\begin{align}
  P_{j}^{{}}=&  \Expect \left[ h_{opt}^{{}}h_{u}^{{}}X_{u}^{{}}X_{u}^{*}h_{u}^{*}h_{opt}^{*} \right] \notag\\
 & =P_{u}^{3}\left| h_{u}^{{}} \right|_{{}}^{4}\frac{\left| h_{i}^{{}} \right|_{{}}^{2}}{\left| h_{d}^{{}} \right|_{{}}^{2}\left( \left| h_{u}^{{}} \right|_{{}}^{2}P_{u}^{{}}+\sigma _{n}^{2} \right)_{{}}^{2}}.
\end{align}

Hence, the average transmit power can be solved by the following equation

\begin{align} \label{eqa:rayleienery2}
P_{u}^{3}\left| h_{u}^{{}} \right|_{{}}^{4}\frac{\left| h_{i}^{{}} \right|_{{}}^{2}}{\left| h_{d}^{{}} \right|_{{}}^{2}\left( \left| h_{u}^{{}} \right|_{{}}^{2}P_{u}^{{}}+\sigma _{n}^{2} \right)_{{}}^{2}}+2P_{u}^{{}}=P
\end{align}

The numerical algorithm of the energy efficiency calculation for case four is summarized as \textbf{Algorithm \ref{algrithm}}.

\begin{algorithm}[h]
    \caption{Energy efficient calculation for fading channels} \label{algrithm}
    \begin{algorithmic}[1]
    \STATE Given  $\beta$, $T_d$, $P$, $\sigma_n^2$, generate the $I$ length channel power vectors  $H_{u2}$, $H_{d2}$, and $H_{i2}$, $R \leftarrow 0$;
    \FOR{$i=1$; $i<I$; $i++$ }
     \STATE  $h_{u2}\leftarrow H_{u2} \left( i \right)$, $h_{d2}\leftarrow H_{d2}\left( i \right)$,  $h_{i2}\leftarrow H_{i2}\left( i \right)$
    \IF{$h_{d2}  > T_d $}
           \STATE Solve equation: $P_{u}^{3}h_{u2}^{2}\frac{ h_{i2}}{ h_{d2}\left(  h_{u2}P_{u}^{{}}+\sigma _{n}^{2} \right)_{{}}^{2}}+2P_{u}^{{}}=P$;
           \STATE $P_d \leftarrow P_u$;
           \IF{$h_{u2} \ge \frac{\beta \sigma_n^2}{P_u }$ }
                \STATE $R \leftarrow R + \log_2\left( 1+\frac{h_{u}P_{u}^{{}}}{\sigma _{n}^{2}} \right) + \log_2 \left(1+ \frac{ h_{d2}P_{d}^{{}}}{\frac{ h_{i2}P_{u}^{{}}\sigma _{n}^{2}}{h_{u2}P_{u}^{{}}+\sigma _{n}^{2}}+\sigma _{n}^{2}}\right)$;
           \ELSE
              \STATE $P_d \leftarrow P/2$, $P_u \leftarrow P/2$;
              \STATE $ R \leftarrow R + \log_2\left( 1+\frac{ h_{u}P_{u}^{{}}}{\sigma _{n}^{2}} \right) + \log_2 \left(1+\frac{ h_{d2}P_{d}^{{}}}{ h_{i2}P_{u}^{{}}+\sigma _{n}^{2}}\right)$;
          \ENDIF
     \ELSE
           \STATE $P_d \leftarrow P/2$, $P_u \leftarrow P/2$;
           \STATE $ R \leftarrow R + \log_2\left( 1+\frac{ h_{u}P_{u}^{{}}}{\sigma _{n}^{2}} \right) + \log_2 \left(1+\frac{ h_{d2}P_{d}^{{}}}{ h_{i2}P_{u}^{{}}+\sigma _{n}^{2}}\right)$;
     \ENDIF
     \ENDFOR
     \STATE Average energy efficiency $\tilde{R} \leftarrow \frac{R}{I}$.
    \end{algorithmic}
\end{algorithm}

\begin{figure}
\begin{center}
\includegraphics[width=0.6\textwidth]{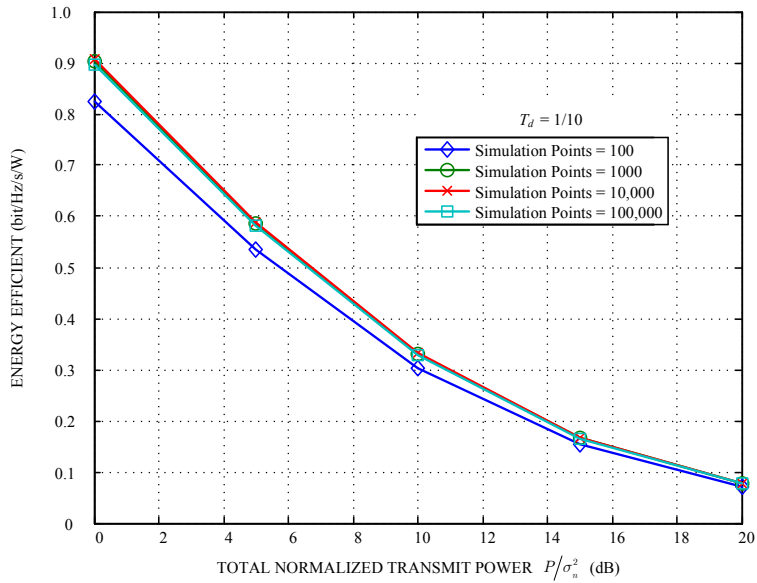}
\end{center}
\centering\caption{Energy efficient vs. the total normalized transmit power $P / \sigma_n^2$ for proposed IUI suppression scheme under different simulation points. The threshold coefficient of the uplink channel power $\beta = 1$.}
\label{f-FadingRayleighEnergyP}
\end{figure}

Fig. \ref{f-FadingRayleighEnergyP} shows the energy efficiency versus the total normalized transmit power $P / \sigma_n^2$ for proposed IUI suppression scheme under different simulation points. The threshold coefficient of the uplink channel power is $\beta = 1$. The  power threshold value of the downlink channel is $T_d = 1/10$. Four cases are simulated. We can see that when the simulation point is larger than $1000$, the Monte Carlo simulations for the energy efficiency are almost stable.

\begin{figure}
\begin{center}
\includegraphics[width=0.6\textwidth]{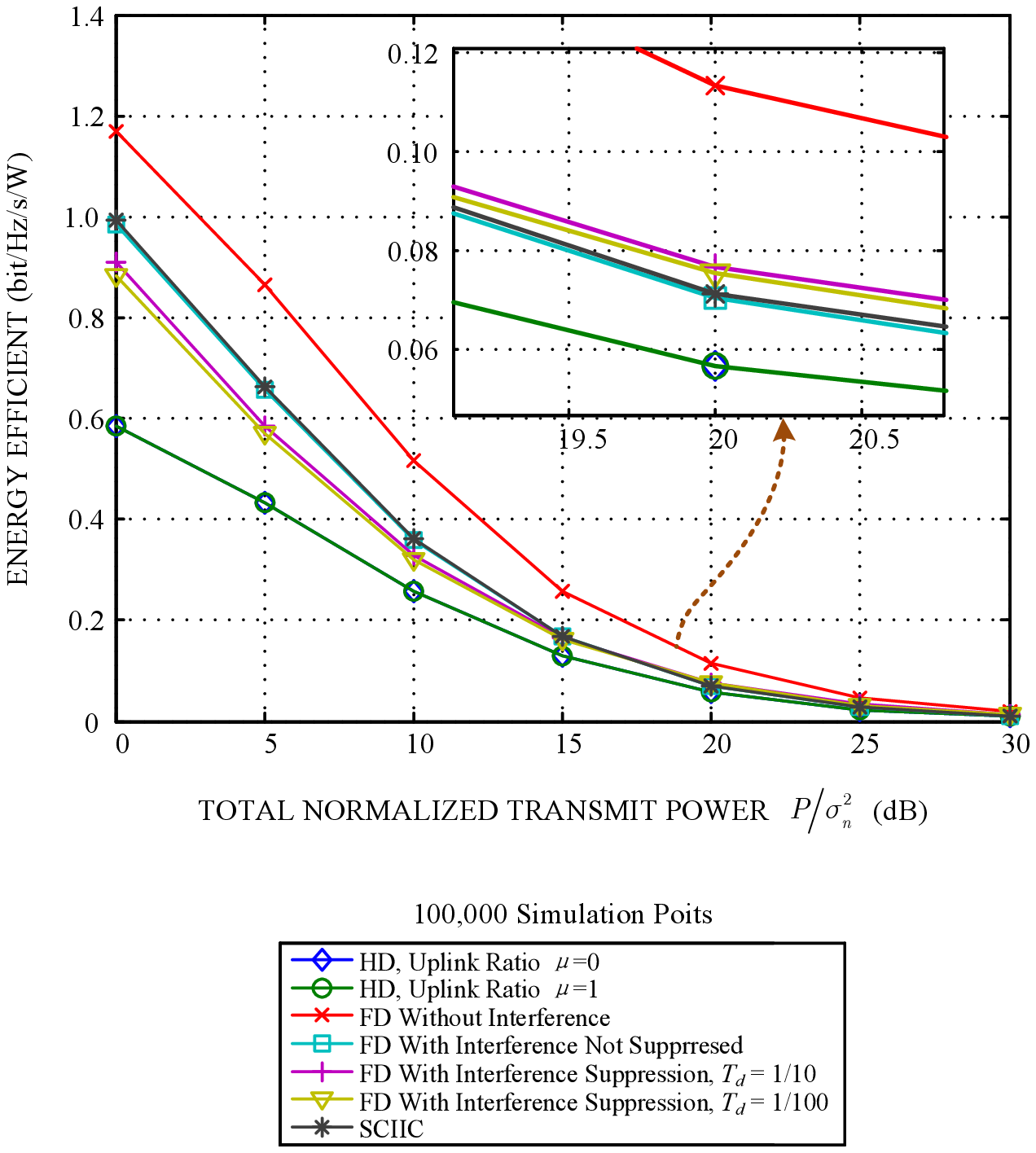}
\end{center}
\centering\caption{Energy efficient vs. the total normalized transmit power $P / \sigma_n^2$. The threshold coefficient of the uplink channel power $\beta = 1$. }
\label{f-FadingRayleighEnergy}
\end{figure}

For the SCIIC scheme, the uplink and downlink transmit power is equal.

Fig. \ref{f-FadingRayleighEnergy} shows the energy efficiency versus the total normalized transmit power $P / \sigma_n^2$ for six cases  besides the SCIIC scheme in \cite{bi2016superposition}.
The threshold coefficient of the uplink channel power $\beta = 1$.
 We can see that the energy efficiency of the FD mode with interference suppressed by the proposed IUI suppression scheme when the $T_d$s are $1/10$ and $1/100$ is higher than the HD mode, but lower than the FD mode with interference not suppressed.
  The energy efficiency of  the FD mode with interference suppressed by the proposed IUI suppression scheme can be improved if we adjust the $T_d$.
 However, too large $T_d$  means low spectral efficiency.
 This result implies that there exist an obvious  trade-off between the  energy efficiency and the spectral efficiency.
 In addition, the proposed scheme performs better than SCIIC when the normalized total transmit power $P/\sigma^2$ is more than about $15$ dB.

 \section{IUI Suppression Scheme for Wideband system}

Here we take OFDM modulation for example.

If we extend the proposed narrowband INI suppression scheme to wideband case, there are two main realizations,  frequency-domain method and time-domain scheme method.

For frequency-domain method, the INI suppression signal is weighted in the frequency-domain. In this case, INI can not be generated and transmitted at the BS until one total OFDM frame has been received. In order to improve the IUI suppression ability, the length of OFDM frame should be designed as short as possible. However, as isillustrated in Section III, to degrade the impact of the delay difference between the INI and INI suppression signals on the IUI suppression ability, the length of OFDM frame should be designed as long as possible. Thus,  this conflict for frequency-domain method makes it worthless in practical application.

For time-domain method, the INI suppression signal is weighted in the time-domain. The optimum weighted coefficient for each subcarrier is translated into time-domain, we call it time-domain INI suppression (TDINIS) filter. Thus, the uplink signal can be transmitted after passing TDINIS filter as soon as it has been received. In this case,  the length of OFDM frame can be set long enough while maintaining the delay difference between the INI and INI suppression signals can be minimized.

In the whole,  the proposed narrowband INI suppression scheme can be extended to the wideband situation  using a TDINIS filter.

\section{Conclusion}
In mixed HD and FD networks, IUI is a key bottleneck especially when the scale of the network becomes larger.
In this paper, one simple IUI suppression scheme is proposed when the BS knows the full state information of  the uplink, the downlink, and the interference channels. We first investigate the effectiveness of the proposed scheme for narrowband case and then extend it to broadband case.

For the narrowband case,  we evaluate the performance of the proposed scheme under practical considerations such as imperfect SI mitigation at the base station, imperfect channel information, delay difference between the IUI and the IUI suppression signals, power control and limited total transmit power.
In practice, the ratio between  delay difference and symbol length, e.g., the length of the OFDM frame, should be designed to be small enough to maintain performance of the proposed approach.
Furthermore,  besides the newest IUI suppression schemes,  we compared the performance of the proposed IUI suppression scheme with the HD mode, and the FD mode with IUI  but not suppressed.
For achievable rate, the proposed approach outperforms other cases or schemes.
For energy efficiency, the proposed approach outperforms other cases or schemes when the SNR of the channels becomes moderately high.
 This implies, the proposed approach prefers high SNR scene such as  small or high density cell, which will be common in the next wireless communications.

 In addition, when applying the proposed IUI suppression method in the FD networks, the users should be paired. How to pair the users?
 One gold rule is according  to the quality of the uplink channel.
  That is to say, the quality of the interference  and the downlink channels is not so important.
  Especially, even in the worst case that the power of the interference channel is strong and the power of the downlink channel is weak, the proposed IUI suppression method can achieve high rate, as long as the quality of the uplink channel is high.
   This is in contrast to the conventional viewpoint that the uplink channel power should be rationally controlled under certain level to obtain optimum sum achievable rate.

\section*{Acknowledgement}
\addcontentsline{toc}{section}{Acknowledgement}

The authors would thank Qingpeng Liang for his helpful advice.  This work was supported by the National Natural Science Foundation of China [grant numbers 61531009, 61471108]; the National Major Projects [grant number 2016ZX03001009]; and the Fundamental Research Funds for the Central Universities.

\bibliographystyle{IEEEtran}
\bibliography{IEEEabrv,WuFeiReferenceT}

\begin{thebibliography}{10}
\providecommand{\url}[1]{#1}
\csname url@samestyle\endcsname
\providecommand{\newblock}{\relax}
\providecommand{\bibinfo}[2]{#2}
\providecommand{\BIBentrySTDinterwordspacing}{\spaceskip=0pt\relax}
\providecommand{\BIBentryALTinterwordstretchfactor}{4}
\providecommand{\BIBentryALTinterwordspacing}{\spaceskip=\fontdimen2\font plus
\BIBentryALTinterwordstretchfactor\fontdimen3\font minus
  \fontdimen4\font\relax}
\providecommand{\BIBforeignlanguage}[2]{{%
\expandafter\ifx\csname l@#1\endcsname\relax
\typeout{** WARNING: IEEEtran.bst: No hyphenation pattern has been}%
\typeout{** loaded for the language `#1'. Using the pattern for}%
\typeout{** the default language instead.}%
\else
\language=\csname l@#1\endcsname
\fi
#2}}
\providecommand{\BIBdecl}{\relax}
\BIBdecl

\bibitem{duarte2010fullduplex}
M.~Duarte and A.~Sabharwal, ``Full-duplex wireless communications using
  off-the-shelf radios: Feasibility and first results,'' in \emph{2010
  Conference Record of the Forty Fourth Asilomar Conference on Signals, Systems
  and Computers}, Nov 2010, pp. 1558--1562.

\bibitem{choi2010achieving}
J.~I. Choi, M.~Jain, K.~Srinivasan, P.~Levis, and S.~Katti, ``Achieving single
  channel, full duplex wireless communication,'' in \emph{Proceedings of the
  sixteenth annual international conference on Mobile computing and
  networking}.\hskip 1em plus 0.5em minus 0.4em\relax ACM, 2010, pp. 1--12.

\bibitem{radunovic2010rethinking}
B.~Radunovic, D.~Gunawardena, P.~Key, A.~Proutiere, N.~Singh, V.~Balan, and
  G.~Dejean, ``Rethinking indoor wireless mesh design: Low power, low
  frequency, full-duplex,'' in \emph{Wireless Mesh Networks (WIMESH 2010), 2010
  Fifth IEEE Workshop on}, June 2010, pp. 1--6.

\bibitem{bharadia2013full}
D.~Bharadia, E.~McMilin, and S.~Katti, ``Full duplex radios,'' \emph{ACM
  SIGCOMM Computer Communication Review}, vol.~43, no.~4, pp. 375--386, 2013.

\bibitem{riihonen2011mitigation}
T.~Riihonen, S.~Werner, and R.~Wichman, ``Mitigation of loopback
  self-interference in full-duplex {MIMO} relays,'' \emph{IEEE Transactions on
  Signal Processing}, vol.~59, no.~12, pp. 5983--5993, Dec 2011.

\bibitem{jain2011practical}
M.~Jain, J.~I. Choi, T.~Kim, D.~Bharadia, S.~Seth, K.~Srinivasan, P.~Levis,
  S.~Katti, and P.~Sinha, ``Practical, real-time, full duplex wireless,'' in
  \emph{Proceedings of the 17th annual international conference on Mobile
  computing and networking}.\hskip 1em plus 0.5em minus 0.4em\relax ACM, 2011,
  pp. 301--312.

\bibitem{duarte2012experimentdriver}
M.~Duarte, C.~Dick, and A.~Sabharwal, ``Experiment-driven characterization of
  full-duplex wireless systems,'' \emph{IEEE Transactions on Wireless
  Communications}, vol.~11, no.~12, pp. 4296--4307, December 2012.

\bibitem{sabbarwal2014inband}
A.~Sabharwal, P.~Schniter, D.~Guo, D.~W. Bliss, S.~Rangarajan, and R.~Wichman,
  ``In-band full-duplex wireless: Challenges and opportunities,'' \emph{IEEE
  Journal on Selected Areas in Communications}, vol.~32, no.~9, pp. 1637--1652,
  Sept 2014.

\bibitem{kim2015asurvey}
D.~Kim, H.~Lee, and D.~Hong, ``A survey of in-band full-duplex transmission:
  From the perspective of {PHY} and {MAC} layers,'' \emph{IEEE Communications
  Surveys Tutorials}, vol.~17, no.~4, pp. 2017--2046, Fourthquarter 2015.

\bibitem{zhang2016fullduplex}
Z.~Zhang, K.~Long, A.~V. Vasilakos, and L.~Hanzo, ``Full-duplex wireless
  communications: Challenges, solutions, and future research directions,''
  \emph{Proceedings of the IEEE}, vol. 104, no.~7, pp. 1369--1409, July 2016.

\bibitem{zhang2015impro}
Y.~Zhang, T.~Liang, and A.~Sun, ``{Improving Physical Layer Security via TAS
  and Full-Duplex Artificial-Noise-Added Receiver},'' \emph{{FREQUENZ}},
  vol.~{69}, no. {7-8}, pp. {357--367}, {JUL} {2015}.

\bibitem{Wu2016nearfield}
\BIBentryALTinterwordspacing
F.~Wu, S.~Li, S.~Shao, and Y.~Tang, ``Near-field self-interference suppression
  with subscriber beamforming in full-duplex communications,''
  \emph{{AEU}-International Journal of Electronics and Communications}, pp.~--,
  2016. [Online]. Available:
  \url{http://www.sciencedirect.com/science/article/pii/S1434841116307257}
\BIBentrySTDinterwordspacing

\bibitem{Syrjl2016analysis}
V.~Syrjala, K.~Yamamoto, and M.~Valkama, ``Analysis and design specifications
  for full-duplex radio transceivers under {RF} oscillator phase noise with
  arbitrary spectral shape,'' \emph{IEEE Transactions on Vehicular Technology},
  vol.~65, no.~8, pp. 6782--6788, Aug 2016.

\bibitem{goyal2013analyzing}
S.~Goyal, P.~Liu, S.~Hua, and S.~Panwar, ``Analyzing a full-duplex cellular
  system,'' in \emph{Information Sciences and Systems (CISS), 2013 47th Annual
  Conference on}, March 2013, pp. 1--6.

\bibitem{goyal2014improving}
S.~Goyal, P.~Liu, S.~Panwar, R.~A. DiFazio, R.~Yang, J.~Li, and E.~Bala,
  ``Improving small cell capacity with common-carrier full duplex radios,'' in
  \emph{2014 IEEE International Conference on Communications (ICC)}, June 2014,
  pp. 4987--4993.

\bibitem{ramirez2013optimal}
D.~Ramirez and B.~Aazhang, ``Optimal routing and power allocation for wireless
  networks with imperfect full-duplex nodes,'' \emph{IEEE Transactions on
  Wireless Communications}, vol.~12, no.~9, pp. 4692--4704, September 2013.

\bibitem{yin2013fullduplex}
B.~Yin, M.~Wu, C.~Studer, J.~R. Cavallaro, and J.~Lilleberg, ``Full-duplex in
  large-scale wireless systems,'' in \emph{2013 Asilomar Conference on Signals,
  Systems and Computers}, Nov 2013, pp. 1623--1627.

\bibitem{shao2014analysis}
S.~Shao, D.~Liu, K.~Deng, Z.~Pan, and Y.~Tang, ``Analysis of carrier
  utilization in full-duplex cellular networks by dividing the co-channel
  interference region,'' \emph{IEEE Communications Letters}, vol.~18, no.~6,
  pp. 1043--1046, June 2014.

\bibitem{nam2015radioresource}
C.~Nam, C.~Joo, and S.~Bahk, ``Radio resource allocation with inter-node
  interference in full-duplex ofdma networks,'' in \emph{2015 IEEE
  International Conference on Communications (ICC)}, June 2015, pp. 3885--3890.

\bibitem{duarte2016interuser}
M.~Duarte, A.~Feki, and S.~Valentin, ``Inter-user interference coordination in
  full-duplex systems based on geographical context information,'' in
  \emph{2016 IEEE International Conference on Communications (ICC)}, May 2016,
  pp. 1--7.

\bibitem{yu2016jointuser}
G.~Yu, D.~Wen, and F.~Qu, ``Joint user scheduling and channel allocation for
  cellular networks with full duplex base stations,'' \emph{IET
  Communications}, vol.~10, no.~5, pp. 479--486, 2016.

\bibitem{sahai2011pushing}
A.~Sahai, G.~Patel, and A.~Sabharwal, ``Pushing the limits of full-duplex:
  Design and real-time implementation,'' \emph{arXiv preprint arXiv:1107.0607},
  2011.

\bibitem{singh2011efficient}
N.~Singh, D.~Gunawardena, A.~Proutiere, B.~Radunovi, H.~V. Balan, and P.~Key,
  ``Efficient and fair {MAC} for wireless networks with self-interference
  cancellation,'' in \emph{Modeling and Optimization in Mobile, Ad Hoc and
  Wireless Networks (WiOpt), 2011 International Symposium on}, May 2011, pp.
  94--101.

\bibitem{kim2013janus}
J.~Y. Kim, O.~Mashayekhi, H.~Qu, M.~Kazandjieva, and P.~Levis, ``Janus: A novel
  mac protocol for full duplex radio,'' \emph{CSTR}, vol.~2, no.~7, p.~23,
  2013.

\bibitem{goyal2013adistributed}
S.~Goyal, P.~Liu, O.~Gurbuz, E.~Erkip, and S.~Panwar, ``A distributed {MAC}
  protocol for full duplex radio,'' in \emph{2013 Asilomar Conference on
  Signals, Systems and Computers}, Nov 2013, pp. 788--792.

\bibitem{choi2015powercontrolled}
W.~Choi, H.~Lim, and A.~Sabharwal, ``Power-controlled medium access control
  protocol for full-duplex {W}i{F}i networks,'' \emph{IEEE Transactions on
  Wireless Communications}, vol.~14, no.~7, pp. 3601--3613, July 2015.

\bibitem{chen2015probabilistic}
S.~Y. Chen, T.~F. Huang, K.~C.~J. Lin, Y.~W.~P. Hong, and A.~Sabharwal,
  ``Probabilistic-based adaptive full-duplex and half-duplex medium access
  control,'' in \emph{2015 IEEE Global Communications Conference (GLOBECOM)},
  Dec 2015, pp. 1--6.

\bibitem{bai2012decode}
J.~Bai and A.~Sabharwal, ``Decode-and-cancel for interference cancellation in a
  three-node full-duplex network,'' in \emph{2012 Conference Record of the
  Forty Sixth Asilomar Conference on Signals, Systems and Computers
  (ASILOMAR)}, Nov 2012, pp. 1285--1289.

\bibitem{bai2013distributed}
------, ``Distributed full-duplex via wireless side-channels: Bounds and
  protocols,'' \emph{IEEE Transactions on Wireless Communications}, vol.~12,
  no.~8, pp. 4162--4173, August 2013.

\bibitem{sahai2013onuplink}
A.~Sahai, S.~Diggavi, and A.~Sabharwal, ``On uplink/downlink full-duplex
  networks,'' in \emph{2013 Asilomar Conference on Signals, Systems and
  Computers}, Nov 2013, pp. 14--18.

\bibitem{sahai2014ondefrees}
------, ``On degrees-of-freedom of full-duplex uplink/downlink channel,'' in
  \emph{Information Theory Workshop (ITW), 2013 IEEE}, Sept 2013, pp. 1--5.

\bibitem{bi2015onrate}
W.~Bi, X.~Su, L.~Xiao, and S.~Zhou, ``On rate region analysis of full-duplex
  cellular system with inter-user interference cancellation,'' in \emph{2015
  IEEE International Conference on Communication Workshop (ICCW)}, June 2015,
  pp. 1166--1171.

\bibitem{bi2016superposition}
------, ``Superposition coding based inter-user interference cancellation in
  full duplex cellular system,'' in \emph{2016 IEEE Wireless Communications and
  Networking Conference}, April 2016, pp. 1--6.

\bibitem{mai2016defrees}
V.~V. Mai, J.~Kim, S.~W. Jeon, S.~W. Choi, B.~Seo, and W.~Y. Shin, ``Degrees of
  freedom of millimeter wave full-duplex systems with partial {CSIT},''
  \emph{IEEE Communications Letters}, vol.~20, no.~5, pp. 1042--1045, May 2016.

\bibitem{chung2001course}
K.~L. Chung, \emph{A course in probability theory}.\hskip 1em plus 0.5em minus
  0.4em\relax Academic press, 2001.

\bibitem{zhang2015investigation}
R.~Zhang, M.~Ma, D.~Li, and B.~Jiao, ``Investigation on dl and ul power control
  in full-duplex systems,'' in \emph{2015 IEEE International Conference on
  Communications (ICC)}, June 2015, pp. 1903--1907.

\bibitem{feng2015joint}
\BIBentryALTinterwordspacing
M.~Feng, S.~Mao, and T.~Jiang, ``Joint duplex mode selection, channel
  allocation, and power control for full-duplex cognitive femtocell networks,''
  \emph{Digital Communications and Networks}, vol.~1, no.~1, pp. 30 -- 44,
  2015. [Online]. Available:
  \url{http://www.sciencedirect.com/science/article/pii/S2352864815000036}
\BIBentrySTDinterwordspacing

\bibitem{imari2016theoretical}
M.~Al-Imari, ``Theoretical analysis of full-duplex system with power control,''
  in \emph{2016 International Symposium on Wireless Communication Systems
  (ISWCS)}, Sept 2016, pp. 461--465.

\bibitem{li2016binary}
R.~Li, Y.~Chen, and Y.~Wu, ``Binary power control for full-duplex networks,''
  in \emph{IEEE 27th Annual International Symposium on Personal, Indoor, and
  Mobile Radio Communications (PIMRC)}, accepted.

\bibitem{mairton2016distributed}
J.~Mairton, S.~Jr, Y.~Xu, G.~Fodor, and C.~Fischione, ``Distributed spectral
  efficiency maximization in full-duplex cellular networks,'' in \emph{2016
  IEEE International Conference on Communications (ICC)}, accepted.

\end{thebibliography}
\addcontentsline{toc}{section}{References}

\end{document}